\documentclass[useAMS,usenatbib]{mn2e}
\input psfig.sty

\def \aj {AJ}
\def \apj {ApJ}
\def \apjl {ApJL}
\def \mnras {MNRAS}
\def \apjs {ApJS}
\def \aap {A\&A}

\def \etal {et~al.~}

\def \spose#1{\hbox  to 0pt{#1\hss}}  
\def \lta{\mathrel{\spose{\lower 3pt\hbox{$\sim$}}\raise  2.0pt\hbox{$<$}}}
\def \gta{\mathrel{\spose{\lower  3pt\hbox{$\sim$}}\raise 2.0pt\hbox{$>$}}}

\def \kmsmpc {\>{\rm km}\,{\rm s}^{-1}\,{\rm Mpc}^{-1}}
\def \kms {\ifmmode  \,\rm km\,s^{-1} \else $\,\rm km\,s^{-1}  $ \fi }
\def \kpc {\ifmmode  {\rm kpc}  \else ${\rm  kpc}$ \fi  }  
\def \hkpc {\ifmmode  {h^{-1}\rm kpc}  \else ${h^{-1}\rm kpc}$ \fi  }  
\def \hMpc {\ifmmode  {h^{-1}\rm Mpc}  \else ${h^{-1}\rm Mpc}$ \fi  }  
\def \Msun {\ifmmode M_{\odot} \else $M_{\odot}$ \fi} 
\def \hMsun {\ifmmode h^{-1}\,\rm M_{\odot} \else $h^{-1}\,\rm M_{\odot}$ \fi}
\def \hhMsun {\ifmmode h^{-2}\,\rm M_{\odot}\else $h^{-2}\,\rm M_{\odot}$ \fi}

\def\LCDM{$\Lambda$CDM }
\def \LCDM {\ifmmode \Lambda{\rm CDM} \else $\Lambda{\rm CDM}$ \fi}
\def \sig8 {\ifmmode \sigma_8 \else $\sigma_8$ \fi} 
\def \OmegaM {\ifmmode \Omega_{\rm M} \else $\Omega_{\rm M}$ \fi} 
\def \Omegab {\ifmmode \Omega_{\rm b} \else $\Omega_{\rm b}$ \fi} 
\def \OmegaL {\ifmmode \Omega_{\rm \Lambda} \else $\Omega_{\rm \Lambda}$\fi} 
\def \Deltavir {\ifmmode \Delta_{\rm vir} \else $\Delta_{\rm vir}$ \fi}
\def \rhocrit {\ifmmode \rho_{\rm crit} \else $\rho_{\rm crit}$ \fi}

\def \rs {\ifmmode r_{\rm s} \else $r_{\rm s}$ \fi} 
\def \rh {\ifmmode r_{\rm h} \else $r_{\rm h}$ \fi} 
\def \Rvir {\ifmmode R_{\rm vir} \else $R_{\rm vir}$ \fi}
\def \Vvir {\ifmmode V_{\rm  vir} \else  $V_{\rm vir}$  \fi} 
\def \Vmax {\ifmmode V_{\rm  max} \else  $V_{\rm max}$  \fi} 
\def \Mvir {\ifmmode M_{\rm  vir} \else $M_{\rm  vir}$ \fi}  

\def \Re {\ifmmode R_{\rm e} \else $R_{\rm e}$ \fi} 
\def \Mstar {\ifmmode M_{\rm star} \else $M_{\rm star}$ \fi} 
\def \Msps {\ifmmode M_{\rm SPS} \else $M_{\rm SPS}$ \fi}

\def \DeltaIMF {\ifmmode \Delta_{\rm IMF} \else $\Delta_{\rm IMF}$ \fi}

\def \gammap {\ifmmode \gamma^{\prime} \else $\gamma^{\prime}$ \fi} 
\def \Rein {\ifmmode R_{\rm Ein} \else $R_{\rm Ein}$ \fi} 

\title[Mass profiles of elliptical galaxies] {The bulge-halo
conspiracy in massive elliptical galaxies: implications for the
stellar initial mass function and halo response to baryonic processes}

\author[Dutton \& Treu]
{Aaron  A. Dutton$^{1}$\thanks{dutton@mpia.de} \& Tommaso Treu$^2$\\
  $^1$Max Planck Institute for Astronomy, K\"onigstuhl 17, D-69117
  Heidelberg, Germany\\
  $^2$Department of Physics, University of California, Santa Barbara, CA 93106, USA\\
}

\begin{document}
             
\date{submitted to MNRAS}
             
\pagerange{\pageref{firstpage}--\pageref{lastpage}}\pubyear{2014}

\maketitle           

\label{firstpage}
             

\begin{abstract}
  Recent studies have shown that massive elliptical galaxies have
  total mass density profiles within an effective radius that can be
  approximated as $\rho_{\rm tot}\propto r^{-\gammap}$, with mean
  slope $\langle \gammap \rangle=2.08\pm0.03$ and scatter
  $\sigma_{\gammap}=0.16\pm0.02$. The small scatter of the slope
  (known as the bulge-halo conspiracy) is not generic in $\Lambda$
  cold dark matter ($\LCDM$) based models and therefore contains
  information about the galaxy formation process. We compute the
  distribution of $\gammap$ for $\LCDM$-based models that reproduce
  the observed correlations between stellar mass, velocity dispersion,
  and effective radius of early-type galaxies in the Sloan Digital Sky
  Survey. The models have a range of stellar initial mass functions
  (IMFs) and dark halo responses to galaxy formation.  The observed
  distribution of $\gammap$ is well reproduced by a model with
  cosmologically motivated but uncontracted dark matter haloes, and a
  Salpeter-type IMF. Other models are on average ruled out by the
  data, even though they may happen in individual cases. Models with
  adiabatic halo contraction (and lighter IMFs) predict too small
  values of $\gammap$. Models with halo expansion, or
  mass-follows-light predict too high values of $\gammap$.  Our study
  shows that the non-homologous structure of massive early-type
  galaxies can be precisely reproduced by \LCDM models if the IMF is
  not universal and if mechanisms such as feedback from active
  galactic nuclei, or dynamical friction, effectively on average
  counterbalance the contraction of the halo expected as a result of
  baryonic cooling.
\end{abstract}


\setcounter{footnote}{1}


\section{Introduction}
\label{sec:intro}

Recent observations combining stellar kinematics with strong
gravitational lensing from the Sloan ACS lens Survey (SLACS; Bolton
\etal 2006, 2008; Auger \etal 2009) have shown the total mass density
profiles of massive elliptical galaxies (i.e. with stellar mass above
$\Mstar\sim10^{11}\Msun$ or stellar velocity dispersion $\sigma>200$
\kms) are very close to isothermal (Treu \& Koopmans 2004; Koopmans
\etal 2006, 2009; Auger \etal 2010b; see also, e.g., Bertin \etal
1994, Franx \etal 1994, Dobke \& King 2006, Gavazzi \etal 2007,
Humphrey \& Buote 2010 for additional probes pointing in the same
direction).  The high precision lensing and dynamical measurements
show that within the effective or half-light radius, the average
logarithmic slope of the mass density profile ($\rho_{\rm tot}\propto
r^{-\gammap}$) is $\langle\gammap\rangle = 2.078\pm{0.027}$ with an
intrinsic scatter of $\sigma_{\gammap}= 0.16\pm0.02$ (Auger \etal
2010b).

The physical origin of this observational result is at present not
fully understood.  We first note that the mass density profiles of
stars and $\Lambda$ cold dark matter ($\LCDM$) haloes are different
from isothermal near the effective radius: If mass-follows-light
(MFL), then for a de Vaucouleurs profile $\gammap \sim 2.3$. In
contrast, if dark matter dominates, then $\gammap \lta 1.5$.  The
observational fact that $\gammap \simeq 2$ implies that both baryons
and dark matter contribute non-negligible fractions to the mass within
the effective radii of massive ellipticals. This is sometimes referred
to as the ``bulge-halo conspiracy'' -- the baryons and dark matter
must ``know'' about each other so that the total mass profile that
results from summing the two non-isothermal components is close to
isothermal. In other words the observation can be qualitatively
reproduced by combining standard \LCDM\ haloes with stellar mass
density profiles, but only with appropriate fine tuning of the stellar
to dark matter ratio, or equivalently of the star formation efficiency
(e.g. Keeton 2001; Gavazzi \etal 2007; Jiang \& Kochanek 2007; Auger
\etal 2010a; Nipoti, Treu \& Bolton 2008).

Remarkably, the bulge-halo conspiracy does not happen at all
scales. For example, the total mass density profile of galaxy clusters
with a massive galaxy at the centre is known to be inconsistent with
isothermal (e.g., Allen \etal 2008; Newman \etal 2013a,b). Since dark
matter haloes are almost scale invariant in \LCDM (except for mild
trends in concentration, e.g. Macci\`o \etal 2008), this difference
suggests that understanding the origin of the bulge-halo conspiracy
requires understanding the relevant baryonic physics. Attention has
therefore turned to cosmological hydrodynamical simulations of galaxy
formation.

Duffy \etal (2010) found that simulations with no or weak feedback can
reproduce the observed $\gammap$, but they overpredict the galaxy
formation efficiencies (see also Naab \etal 2007).  Strong feedback is
needed to match the observed baryon fractions, but the resulting
galaxies have $\gammap$ that is too small (i.e. the galaxies are dark
matter dominated).  The simulations used by Duffy \etal (2010) were
stopped at $z=2$, so their conclusions are only valid if $\gammap$
does not evolve. Johansson \etal (2012) ran nine high resolution
``zoom-in'' simulations of $\sim 10^{12}\Msun$ dark matter haloes down
to $z=0$ (see also Remus \etal 2013). These simulations have $\gammap$
in good agreement with observations at $z\sim 0$, and also find
$\gammap$ evolves from $\gammap\sim 3$ at high redshifts to
$\gammap\sim 2$ below $z\sim 1$. However, these simulations (which do
not have strong feedback) overpredict the galaxy formation
efficiencies by a factor of $\sim 2-3$, even assuming a Salpeter
(1955) stellar initial mass function (IMF).
More recently Dubois \etal (2013) simulated six haloes in the mass
range $0.4 - 8\times 10^{13}\Msun$ both with and without feedback from
active galactic nuclei (AGN). At $z=0$ the simulations without AGN
feedback resulted in too many baryons and overpredict the mass density
slopes ($\gammap\sim 2.3$), while the simulations with AGN feedback
reproduce the galaxy formation efficiencies (assuming heavy IMFs in
more massive galaxies), but under predict the mass density slopes
($\gammap\sim 1.9$).
{\it Thus at present, none of the cosmological simulations in the
literature can simultaneously match the observed $\gammap$ and galaxy
formation efficiencies at $z\sim 0$.}

In this paper we aim to shed light on this issue by performing a
quantitative comparison of flexible \LCDM-inspired models to the
observed distribution of $\gammap$ in lens galaxies. First, we
investigate how dark matter fraction is required in order for models
to reproduce the observed distribution of $\gammap$. If a model has
the right (i.e., observed) amount (and distribution of baryons) are
the theoretical properties of \LCDM haloes sufficient to explain the
observed properties of $\gammap$?
A second question we address is to what extent $\gammap$ can help to
constrain the dark matter fractions inside an effective radius, and
hence to constrain the stellar IMF and dark halo response to galaxy
formation.

Observationally, a common way of measuring the dark matter fraction
within an effective radius is achieved by ``subtracting'' the stellar
mass from the total mass. Total masses are relatively straight forward
to measure (either through dynamics and/or strong lensing), but
stellar masses from spectral energy distributions are uncertain by a
factor of $\sim 3$ due to the unknown form of the IMF (e.g. Bell \etal
2003).

Star counts in stellar clusters are consistent with a universal IMF
{\it within the Milky Way} (e.g., Bastian, Cover, Meyer
2010). Dynamical mass-to-light ratios of extragalactic spiral galaxies
are also consistent with a Milky Way IMF (e.g., Bell \& de Jong 2001),
and inconsistent with the heavier\footnote{By heavier we mean that the
  IMF results in higher stellar mass-to-light ratios in old stellar
  populations. This could result from either an excess of low-mass
  stars (from a bottom heavy IMF), or from an excess of stellar
  remnants (from a top heavy IMF)} Salpeter (1955) IMF. Dynamical
mass-to-light ratios of galaxies in the SAURON survey argue against a
universal Salpeter IMF in early-type galaxies (Cappellari \etal 2006;
Tortora \etal 2012), but they do not rule out a Salpeter-type IMF in
all early-type galaxies. In fact, Salpeter-type IMFs are preferred in
massive elliptical galaxies, even accounting for standard dark matter
haloes (Treu \etal 2010; Dutton \etal 2011). Furthermore, there are
several lines of evidence that {\it require} massive elliptical
galaxies and even massive spiral bulges to have IMFs significantly
heavier than that found in the Milky Way (e.g., Auger \etal 2010a; van
Dokkum \& Conroy 2010; Conroy \& van Dokkum 2012; Cappellari \etal
2013; Dutton \etal 2013a,b).

Theoretically, the dark matter fraction within the effective radius
depends on the efficiency of galaxy formation, the structure of
``pristine'' dark matter haloes, and the response of the dark matter
halo to galaxy formation. Galaxy formation efficiencies are hard to
predict theoretically, but can be constrained through observations of
weak lensing and satellite kinematics (e.g., Mandelbaum \etal 2006;
More \etal 2011).  In the framework of \LCDM the pristine structure of
dark matter haloes is known (for a given set of cosmological
parameters), but the response of the dark matter halo is not well
constrained by theory. Gas cooling or gas rich (dissipational) mergers
are expected to make the dark halo contract (Blumenthal \etal 1986;
Gnedin \etal 2004). Mass outflows (feedback) or gas poor
(dissipationless) mergers are expected to make the dark halo expand
(e.g., Read \& Gilmore 2005; Governato \etal 2010; Pontzen \&
Governato 2012; Macci\`o \etal 2012; El-Zant \etal 2001; Nipoti \etal
2004; Jardel \& Sellwood 2009; Johansson \etal 2009; Teyssier
  \etal 2011). The relative importance of each of these processes
will determine the response of the dark matter halo to galaxy
formation.

In practice, we address the two questions by using \LCDM based models
(from Dutton \etal 2013b) that are constructed to reproduce a number
of scaling relations of early-type galaxies including: velocity
dispersion versus stellar mass; half-light size versus stellar mass
and dark halo mass versus stellar mass. The main unknowns in these
models are the normalization of the stellar mass (and therefore the
stellar IMF) and the response of the dark matter halo to galaxy
formation.

This paper is organized as follows: our definition of $\gammap$ is
given in \S2. The constrained galaxy mass models are briefly discussed
in \S3. Results are presented in \S4. The method used to measure
ensemble average mass density slopes is discussed in \S5. A brief
discussion is given in \S6, followed by conclusions in \S7.
We adopt a \LCDM cosmology with $\OmegaL=0.7$,
$\OmegaM=0.3$ and $H_0=70 \kmsmpc$

\section{Definition of mass density slope}

There are many different definitions for the mass density slope of
galaxies used in the literature. We have chosen our definition
primarily to enable a meaningful comparison to the mass density slopes
derived from joint strong lensing and dynamics analyses (e.g.,
Koopmans \etal 2009). In addition, our definition is simple to compute
and can be applied to all type of galaxies (spirals as well as
ellipticals).

We adopt the mass weighted slope within the effective radius (which,
we show below is equivalent to the mass slope at the effective radius)
as heuristically strong lensing and stellar dynamics measure the total
{\it mass} (rather than density) within different radii. These radii
are typically close to the effective radius, hence a joint strong
lensing and dynamics analysis measures the slope of the total mass
profile near the effective radius.  

For an arbitrary density profile, $\rho(r)\propto r^{-\gamma(r)}$, the
local logarithmic slope of the density profile is $d\log\rho/d\log r
\equiv -\gamma(r)$. We follow the convention of Koopmans \etal (2009)
that $\gamma > 0$. Thus the mass weighted slope of the density profile
within radius $r$ is given by
\begin{eqnarray}
\label{eq:gamma1}
\gammap(r) = \frac{1}{M(r)} \int_0^r -\gamma(x) 4\pi x^2\rho(x)\,dx
= 3-\frac{4 \pi r^3\rho(r)}{M(r)}. 
\end{eqnarray}
This can be expressed in terms of the local logarithmic slopes of the
mass, $M(r)$, and circular velocity, $V(r)$ profiles:
\begin{equation}
\label{eq:gamma2}
\gammap(r) = 3-\frac{d\log M}{d \log r} = 2-2\frac{d\log V}{d \log r}. 
\end{equation}
In the case of a power-law density profile $\gammap(r)=\gamma$.  But
in general, i.e., for a non power-law density profile, $\gammap(r) \ne
\gamma(r)$. Since galaxies have mass density profiles with slopes that
decrease (i.e., become more negative) with increasing radius, the mass
weighted slope is less than the local slope: $\gammap(r) < \gamma(r)$.
For example, for an NFW profile (which has an inner density slope of
$-1$ and outer slope of $-3$) at the scale radius $\gamma(r_{\rm
  s})=2$ whereas $\gammap(r_{\rm s})\simeq 1.7$, and for a Hernquist
profile (inner slope $-1$, outer slope $-4$) at the effective radius
$\gamma(R_{\rm e})\simeq 2.9$ whereas $\gammap(R_{\rm e})\simeq2.3$.
In the following we will use the shorthand notation $\gammap$ and to
indicate the mass weighted average slope within the effective radius.


\section{Mass models of early-type galaxies with $\Lambda$CDM haloes.}
\label{sec:mm}
This section gives a brief overview of the mass models we construct to
reproduce the observed structural and dynamical scaling relations of
early-type galaxies in the Sloan Digital Sky Survey (SDSS; York \etal
2000; Abazajian \etal 2009). A more detailed discussion is given by
Dutton \etal (2013b).  Our mass models consist of spherical
distributions of stars and dark matter. The stars are modelled as the
sum of two S\'ersic profiles (with $n=1$ and $n=4$), while the dark
matter is an NFW (Navarro, Frenk, \& White 1997) profile with
``pristine'' concentration parameters from Macci\`o \etal (2008) and
modified by halo contraction/expansion following Dutton \etal (2007).

High resolution cosmological simulations have shown that at small
radii ($\sim 1\%$ of the virial radius) dark matter haloes have
steeper density profiles than the NFW formula. A number of studies
have shown that the Einasto (1965) profile ($d\ln\rho/d\ln r \propto
r^{\alpha}$), provides, in general, a better description of CDM haloes
than the NFW profile (e.g., Navarro \etal 2004, 2010; Merritt \etal
2005, 2006; Stadel \etal 2009; Reed \etal 2011).  Reed \etal (2011)
show that the variation in density profiles of massive dark matter
haloes are fully explained by an Einasto profile with $\alpha=0.19$
together with the usual scatter in the halo concentration parameter.
To determine whether our results are sensitive to the difference
between NFW and Einasto haloes at small radii we have re-computed our
uncontracted models using the Einasto profile, and kept all other
parameters the same as in the original NFW model.  The increased dark
matter density at small radii from the Einasto profile results in just
a 1\% increase in aperture velocity dispersion, 2\% decrease in mass
density slope ($\gammap$) and a 6\% increase in the scatter in
$\gammap$.  Thus in what follows we predominantly focus on results
with NFW haloes.

The stellar profiles of the models are constructed to reproduce the
size-mass relation of early-type galaxies from Dutton \etal (2013b)
which uses sizes from Simard \etal (2011), and stellar masses from the
MPA/JHU group\footnote{ Available at
  http://www.mpa-garching.mpg.de/SDSS/DR7/} which assume a Chabrier
(2003) IMF.  The halo masses are chosen to follow the observed
relation between stellar mass and halo mass for early-type galaxies
from Dutton \etal (2010) which combines results from weak lensing
(e.g., Schulz \etal 2010) and satellite kinematics (e.g., More \etal
2011).  The models have three sources of scatter: scatter in galaxy
size at fixed stellar mass (constrained by observations: Dutton \etal
2013b); scatter in dark halo mass at fixed stellar mass (constrained by
observations: More \etal 2011); and scatter in dark halo concentration
at fixed dark halo mass (constrained by theory: Macci\`o \etal 2008).

The velocity dispersion versus stellar mass (a.k.a. Faber \& Jackson
1976) relation is used to constrain the allowed combinations of
stellar IMF and dark halo response in these models.  We consider five
different halo responses ranging from standard adiabatic contraction
(Blumenthal \etal 1986) to unmodified NFW haloes to MFL (i.e., maximum
expansion).  In order to reproduce the slope and zero-point of
Faber-Jackson relation, all these models require ``heavier'' IMFs in
more massive galaxies, and (trivially) ``heavier'' IMFs in models with
stronger halo expansion.  These set of models are physically realistic
(at least in terms of reproducing the scaling relations) while having
a wide range of dark matter fractions and inner dark matter halo
density profiles.

The correlation between the scatter in the velocity-mass relation with
the scatter on the size-mass relation -- equivalent to the tilt of the
Fundamental Plane (FP; Dressler \etal 1987; Djorgovski \& Davis 1987;
Ciotti \etal 1996) -- was used by Dutton \etal (2013b) as an
additional constraint to distinguish between these models.  For
galaxies in the mass range $10^{10} \lta M_{\rm SPS} \lta
10^{11}\Msun$, only models in which MFL were able to match this
additional constraint. For the most massive galaxies (and those which
are relevant for this paper), models with unmodified NFW haloes were
favoured over MFL and adiabatic contraction models.


\section{Results}
\label{sec:results}

\begin{figure}
\centerline{
\psfig{figure=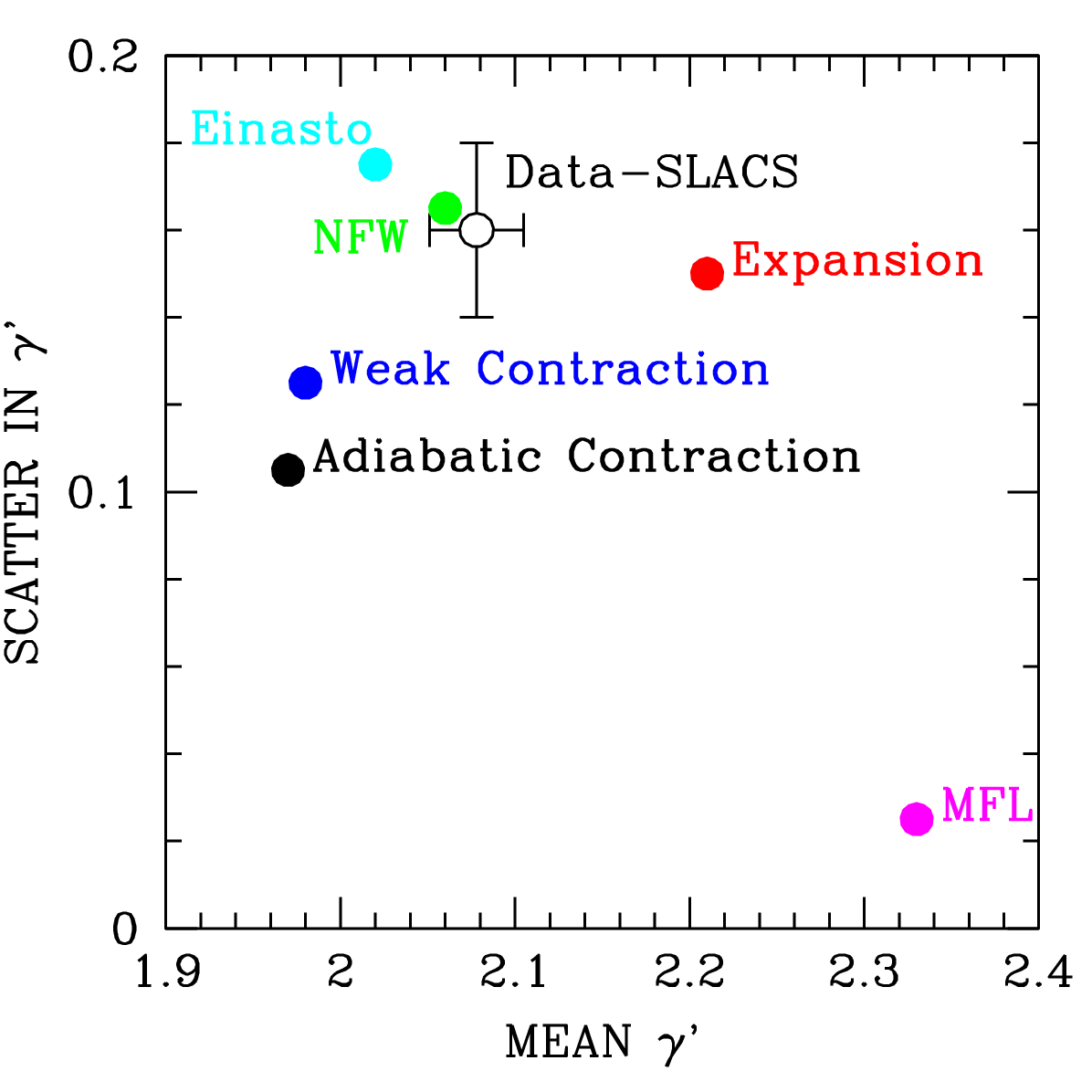,width=0.45\textwidth}
}
\caption{Mean and scatter of the mass weighted density slope within an
  effective radius, $\gammap$, for models (from Dutton \etal 2013b)
  and data (from SLACS -- Auger \etal 2010b). The mean and scatter of
  $\gammap$ are only jointly reproduced in our model with uncontracted
  cosmologically motivated NFW haloes and a heavy Salpeter-type IMF
  (green point). Models with expansion (red point) or mass follows
  light (MFL - magenta point) overpredict the mean $\gammap$, while
  models with halo contraction (blue and black) under predict both the
  mean and scatter. A model with unmodified Einasto dark matter
    haloes (cyan point) yields similar results as the NFW model. }
\label{fig:gamma-mean-sigma}
\end{figure}

\begin{figure*}
\centerline{ 
\psfig{figure=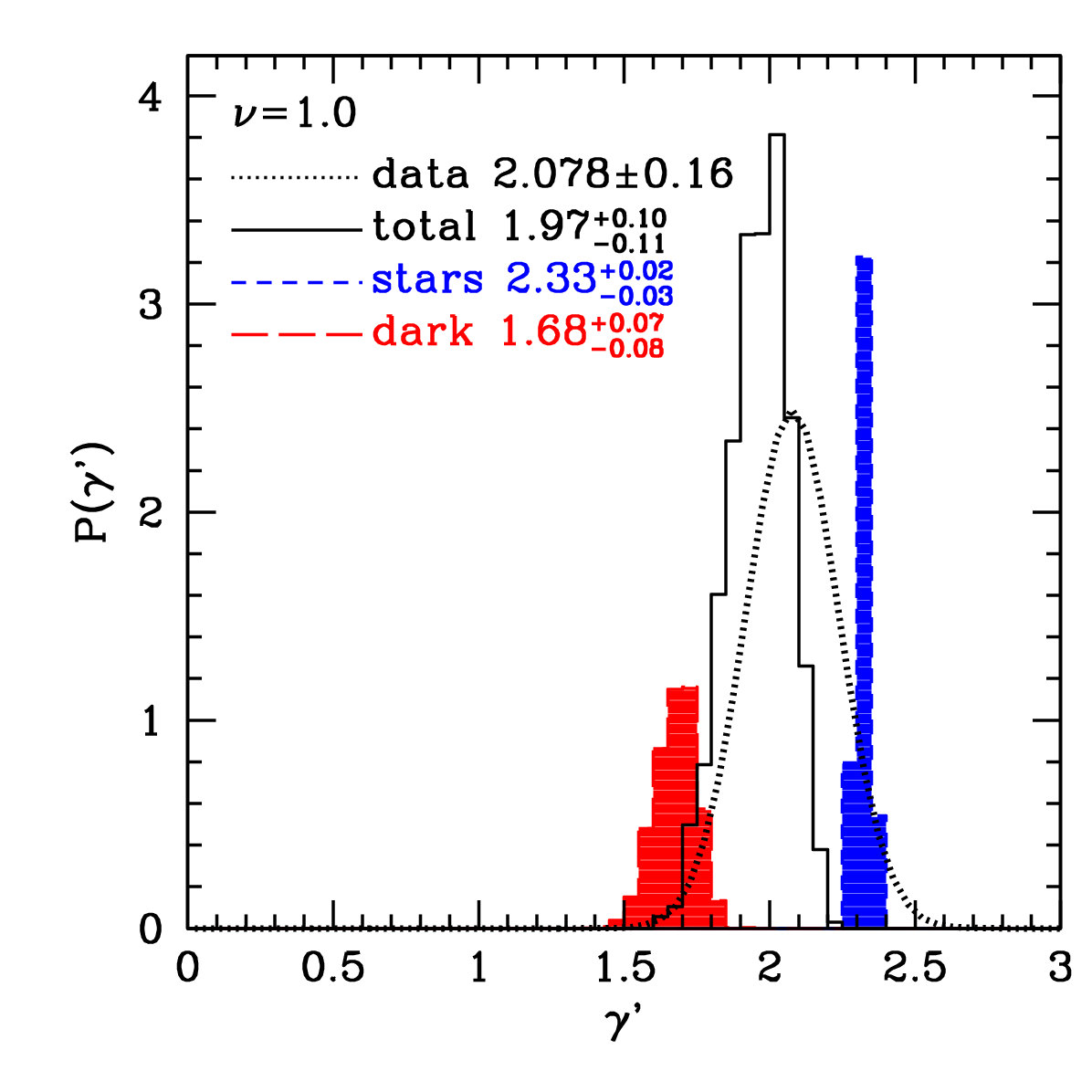,width=0.34\textwidth}
\psfig{figure=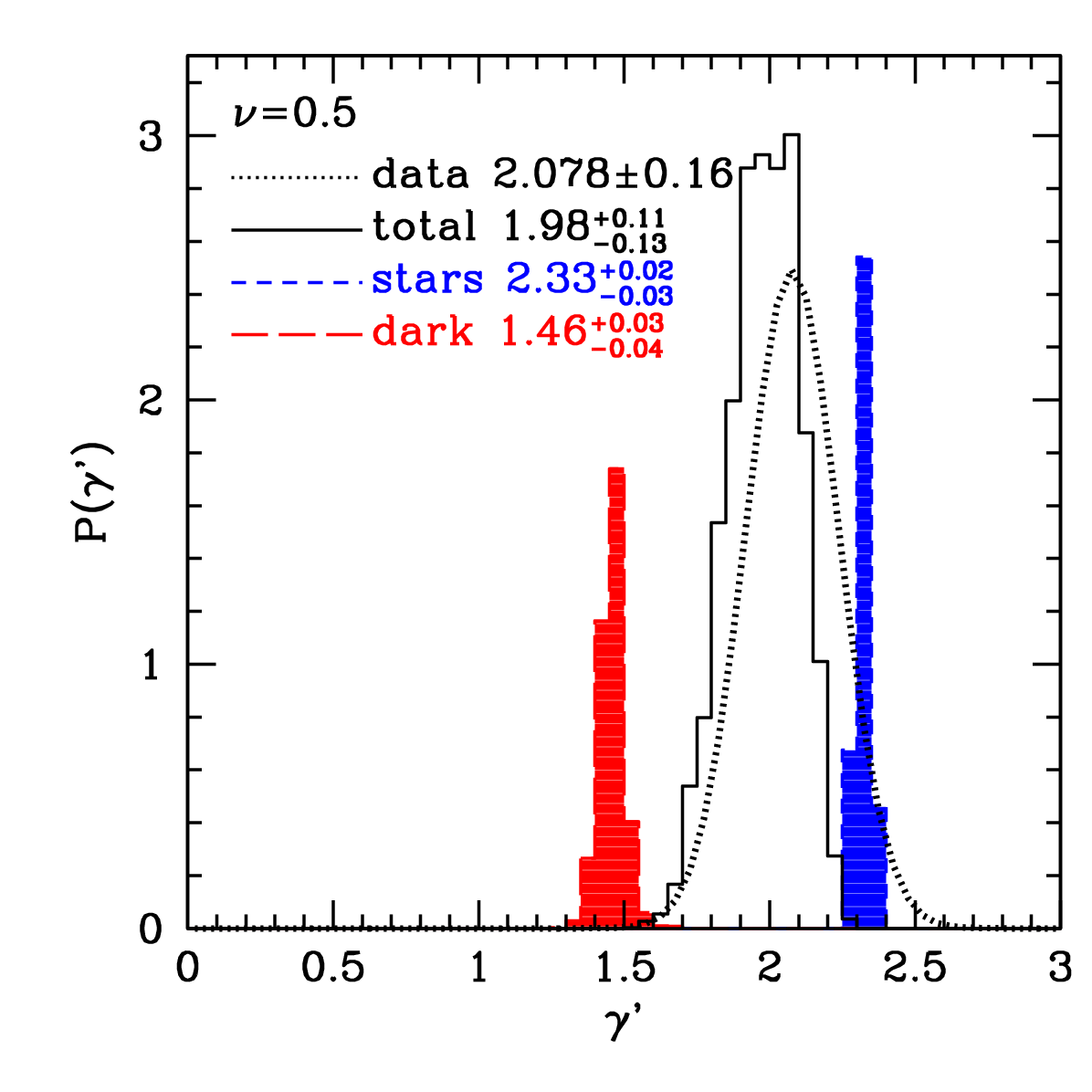,width=0.34\textwidth} 
} 
\centerline{ 
\psfig{figure=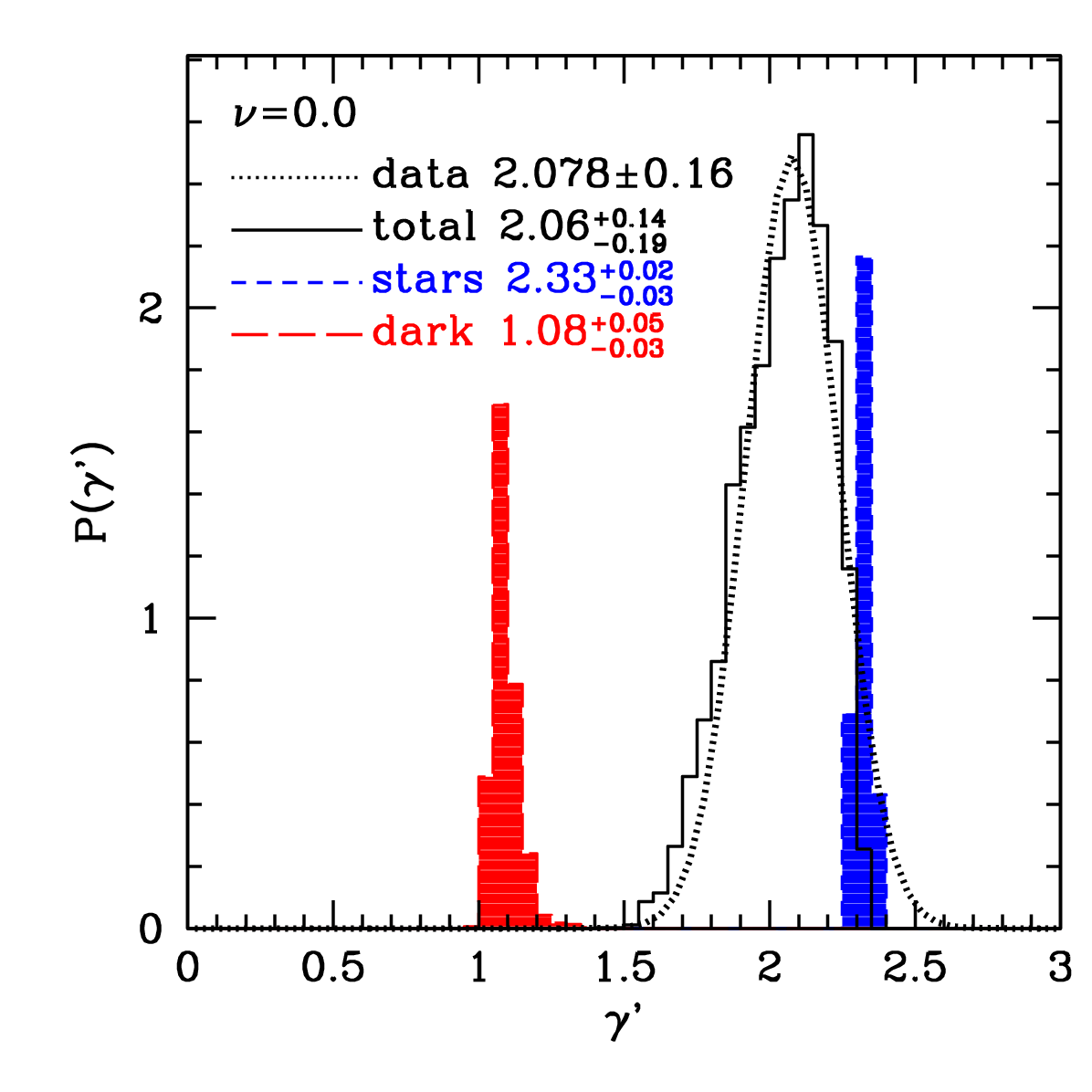,width=0.34\textwidth}
\psfig{figure=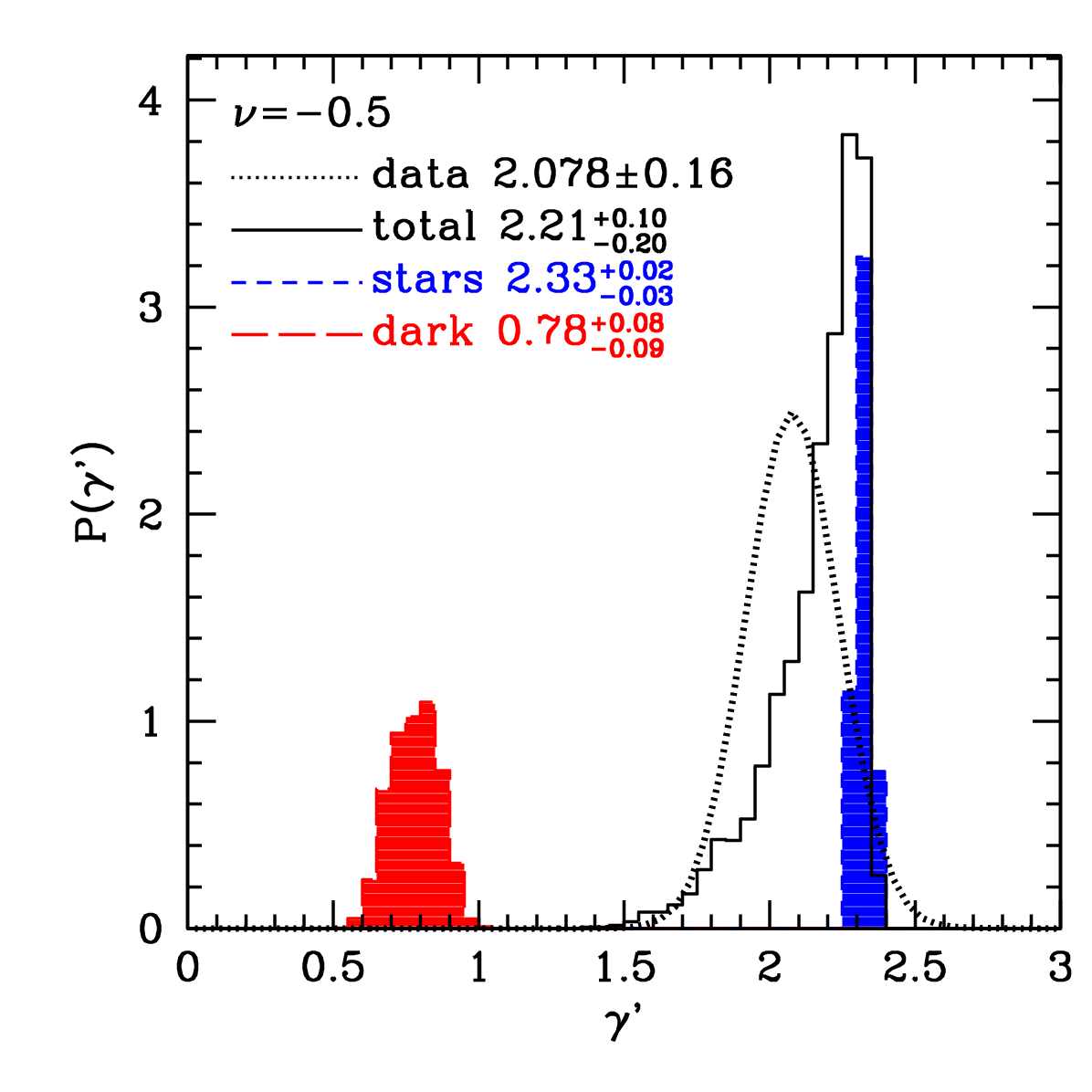,width=0.34\textwidth} 
}
\caption{Distribution of mass-weighted density slope within an
  effective radius, $\gammap$, for our models compared with the
  observed (intrinsic distribution) from Auger \etal (2010b, black
  dotted line). The histograms show $\gammap$ for the total mass
  (black, solid), stellar mass (blue, short dashed), and dark matter
  (red, long dashed).  Each panel has a model with a different dark
  halo response: adiabatic contraction, $\nu=1.0$ (upper left); mild
  contraction $\nu=0.5$ (upper right); no contraction $\nu=0.0$ (lower
  left); and expansion $\nu=-0.5$( lower right). The model with
  mass-follows-light is equivalent to that of the stellar
  component. Only the model with uncontracted cosmologically motivated
  dark matter haloes and a heavy IMF reproduces the observations.}
\label{fig:gammae-hist}
\end{figure*}

In order to make a fairer comparison between our models and data from
SLACS we consider model galaxies that match the distribution of
velocity dispersions for the SLACS lenses. Specifically, we select
model galaxies with a log-normal filter on velocity dispersion: mean
$\log(\sigma/[\kms])=2.40$ and standard deviation $0.07$.  With this
selection the average offset between the model stellar masses and
those obtained assuming a Chabrier IMF are $\Delta_{\rm
  IMF}=\log_{10}(\Mstar/\Msps) = (0.08,0.20, 0.27, 0.31, 0.33)$ for
halo responses of $\nu=(1.0,0.5,0.0,-0.5, {\rm MFL})$, respectively
\footnote{For easy comparison with previous studies $\Delta_{\rm
    IMF}=\log \alpha$, where $\alpha$ is the so-called IMF mismatch
  parameter, e.g., Treu \etal (2010), Dutton \etal (2013a).}. For
reference, a Salpeter IMF corresponds to $\DeltaIMF\simeq 0.23$, and
thus, models with uncontracted dark matter haloes ($\nu=0.0$) and
expansion ($\nu=-0.5$) have IMFs heavier than Salpeter.

The observed $\gammap$ are not strictly measured within any uniform
aperture such as one effective radius, $\Re$. The aperture depends on
the Einstein radius and the physical aperture of the SDSS fibre. The
former depends on the velocity dispersion of the lens and the
redshifts of the lens and source, while the latter just depends on the
redshift of the lens. However, for SLACS lenses the aperture is
typically between one and one-half an effective radii (e.g., Koopmans
\etal 2009). To test whether $\gammap$ depends on the radius within
which it is measured we have compared $\gammap$ within $\Re$ and
$\Re/2$. In general $\gammap(\Re/2) < \gammap(\Re)$, but the changes
are small. The changes in the median \gammap for our five models are
as follows: $\gammap(\Re)-\gammap(\Re/2) = (0.028, 0.022, 0.001, 0.070,
0.141)$ for halo responses of $\nu=(1.0,0.5,0.0,-0.5,{\rm MFL})$.
Thus, the mass density slopes varies between one and one-half an
effective radius by much less than the intrinsic scatter, except for
the MFL model, which we will show below is a poor match to the data
for other reasons (and is well known to be a poor match to massive
elliptical galaxies anyway). For simplicity, in what follows we
measure the mass density slopes within one effective radius, with no
significant loss of precision.

Our main results are summarized in Fig.~\ref{fig:gamma-mean-sigma},
which shows the scatter in $\gammap$ versus the mean of $\gammap$ for
our five models and observations from Auger \etal (2010b).  The
distributions of $\gammap$ for the four models with dark matter haloes
are shown in Fig.~\ref{fig:gammae-hist}. The model with MFL is only
comprised of stars and therefore has $\gammap=2.33^{+0.02}_{-0.03}$.

\subsection{Average mass density slopes}

All of the models broadly reproduce the close to isothermal density
profiles that are observed, with average slopes in the range $1.8 \lta
\gammap \lta 2.4$. However the model with uncontracted NFW haloes (and
a slightly heavier than Salpeter IMF) provides the best match to the
observed average $\gammap$ (Fig.~\ref{fig:gamma-mean-sigma} and yellow
shaded region in Fig.~\ref{fig:gammae-hist}). This is in good
agreement with the conclusions of Auger \etal (2010a) who favoured
uncontracted haloes over any type of halo contraction (they did not
consider expansion or MFL), and IMFs significantly heavier than
Chabrier. It is also in excellent agreement with the, {\it completely
  independent}, constraints from the tilt of the fundamental plane
from Dutton \etal (2013b).

\begin{figure}
\centerline{
\psfig{figure=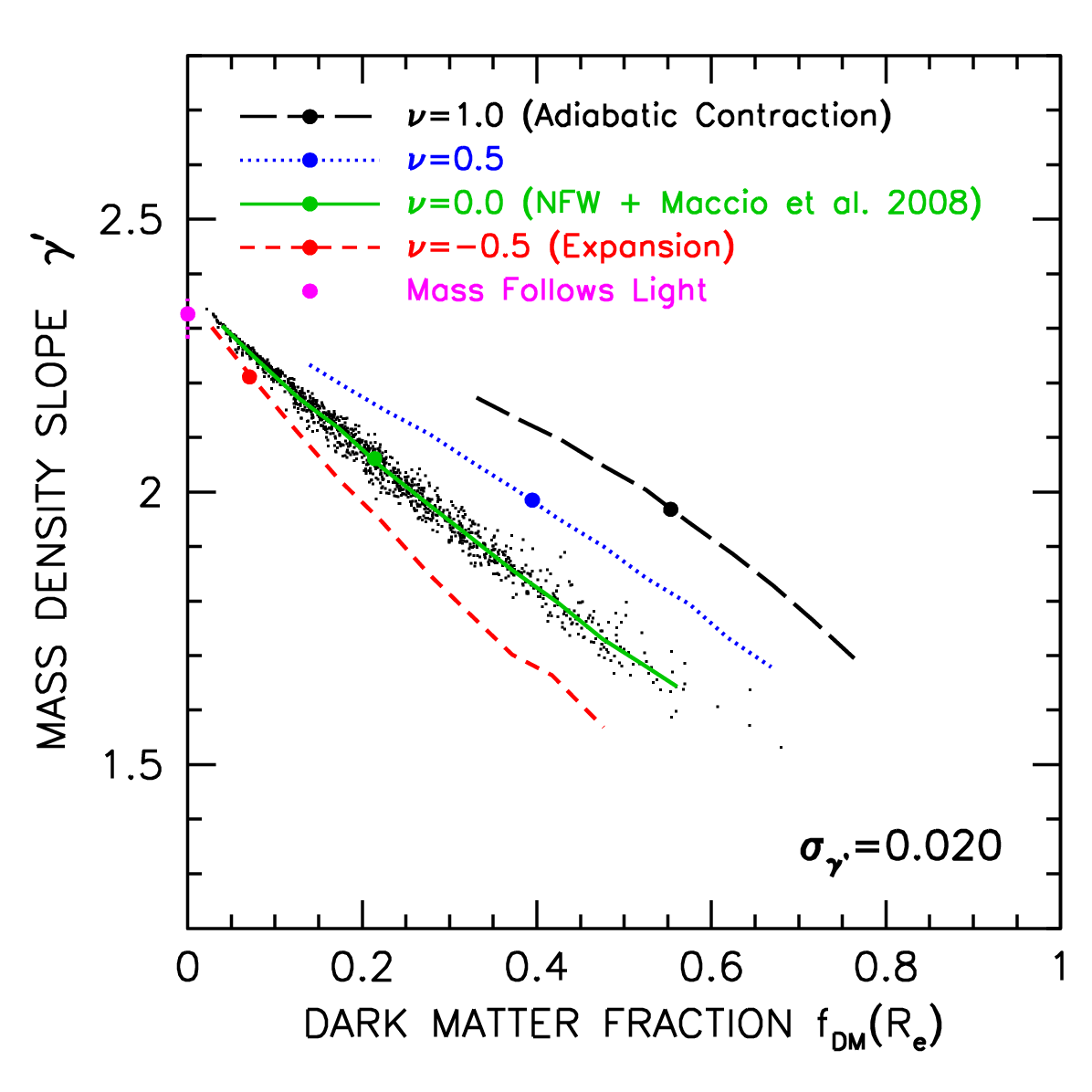,width=0.45\textwidth}
}
\caption{Correlation between mass density slope, $\gammap$, and dark
  matter fraction within an effective radius, $f_{\rm DM}$, for our
  models. The coloured lines show the median mass density slope as a
  function of the (spherical) dark matter fraction, while the dots
  show the median values. For a given halo response the mass density
  slope is determined by the dark matter fraction.}
\label{fig:gammae-fDM}
\end{figure}

\subsection{Scatter in mass density slopes}

Most importantly however, our models also reproduce the observed
scatter of the distribution, i.e. the tightness of the bulge-halo
conspiracy.  The distribution of $\gammap$ in the models is roughly
Gaussian (although slightly skewed) with scatter smaller or equal to
the observed intrinsic scatter of $0.16\pm0.02$. As with the mean
$\gammap$, the model with uncontracted NFW haloes best matches the
scatter with $\gammap=2.06^{+0.14}_{-0.19}$.

Given that there is significant overlap between the observed
distribution of $\gammap$ and all of our models, it is entirely
possible that the full range of halo responses occurs in real
galaxies. Indeed, there are already observational hints that this is
the case. While on average a Salpeter-type IMF favours uncontracted
NFW haloes (e.g., Treu \etal 2010; Auger \etal 2010a; Dutton \etal
2011, 2013b), the detailed study of a massive elliptical, enabled by
the presence of a double Einstein Ring, favours a Salpeter-type IMF and
a contracted halo (Sonnenfeld \etal 2012). From a theoretical
standpoint, this could be connected to the specific merger history of
each galaxy (e.g., Nipoti \etal 2012 and references therein) or
perhaps to the mode of star formation and resulting IMF (Hopkins
2013). Below, we will explore whether \gammap correlates with other
observables. This could potentially provide some clues as to the
origin of each galaxy.

\subsection{Mass density slopes of stars and dark matter}
In Fig.~\ref{fig:gammae-hist}, the red histograms show the
mass-weighted density slopes within $\Re$ of the dark matter haloes,
$\gammap_{\rm DM}$. These range from $\simeq 1.7$ for adiabatic
contraction, to $\simeq 0.8$ for expanded haloes.  The blue histogram
shows the mass-weighted density slopes within $\Re$ of the stars,
these have $\gammap_{\rm star}\simeq 2.33\pm0.02$.

The relative contribution of the stars and dark matter to the total
$\gammap$ is given by the dark matter fraction within an
effective radius, $f_{\rm DM}(\Re)$. As expected, for a given halo
response model $\gammap$ is strongly anti-correlated with the dark
matter fraction (Fig.~\ref{fig:gammae-fDM}).  The lines show the
median $\gammap$ as a function of dark matter fraction, while the dots
show the median $\gammap$ and $f_{\rm DM}(\Re)$. While for an
arbitrary halo response $\gammap$ does not uniquely predict the dark
matter fraction, it does place some useful limits. For example, if
$\gammap > 2$, then $f_{\rm DM}(\Re) < 0.2$.

The average dark matter fractions vary from $\simeq 0.55$ for
adiabatic contracted NFW haloes (black dot) to $\simeq 0.10$ for
expanded NFW haloes (red dot).  For NFW haloes (green dot), the
average dark matter fraction is just $\simeq 20\%$, but this is enough
to lower the mean $\gammap$ by 0.27 and increase the scatter in
$\gammap$ by 0.14 relative to the MFL model.

Note that in our model the variation in dark matter fraction within
the effective radius is driven by all three sources of scatter: the
global stellar to virial mass ratio; the concentration parameter of
the halo; and the effective radius of the galaxy. The former two
result in scatter in the dark matter mass within fixed radius, while
the latter changes the physical radius where the dark matter fraction
is measured (as well as changing the dark matter mass through the
differential halo response).

\subsection{What does $\gammap$ correlate with?}

Previous observational studies have shown that $\gammap$ is largely
uncorrelated with many physical properties of galaxies, and does not
evolve significantly since $z\sim1$ (e.g., Koopmans \etal 2009; Auger
\etal 2010b; see however Ruff \etal 2011 and Bolton \etal 2012 for
tentative detections of a mild evolutionary trend).  The most
significant correlation found by Auger \etal (2010b) was between
$\gammap$ and total surface density, with denser galaxies having
higher $\gammap$. Such a correlation is expected because denser
galaxies have either higher density stellar components (which
increases the contribution of the stars to $\gammap$) and/or stronger
halo contraction (which results in higher $\gammap_{\rm DM}$). This
correlation might provide an important clue to understanding the origin
of the small variations from galaxy to galaxy. The central density
could be a by-product of the epoch of formation of the initial core and
successive merger history and perhaps be related to the conditions of
the gas that created the bulk of the old stars.

\begin{figure*}
\centerline{
\psfig{figure=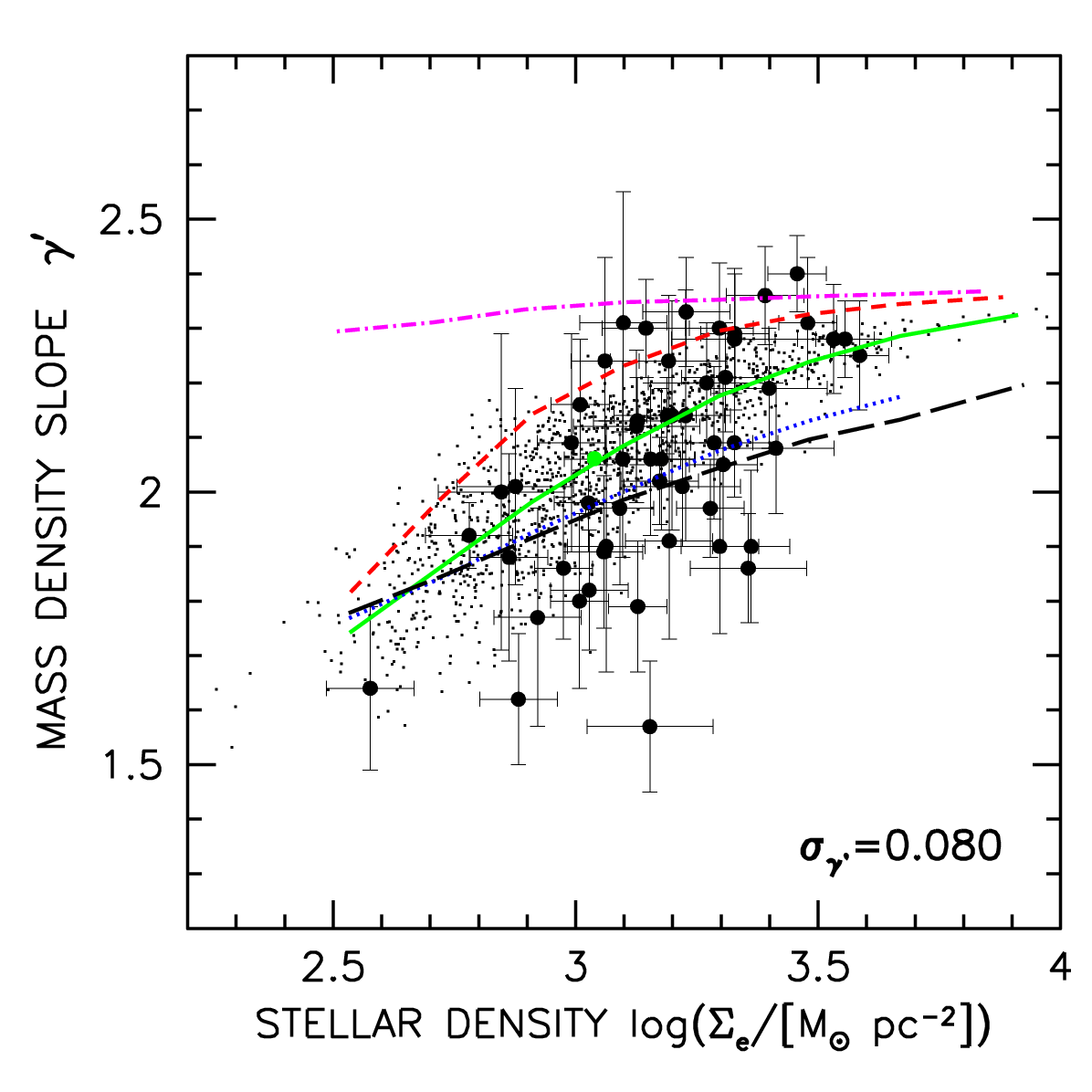,width=0.34\textwidth}
\psfig{figure=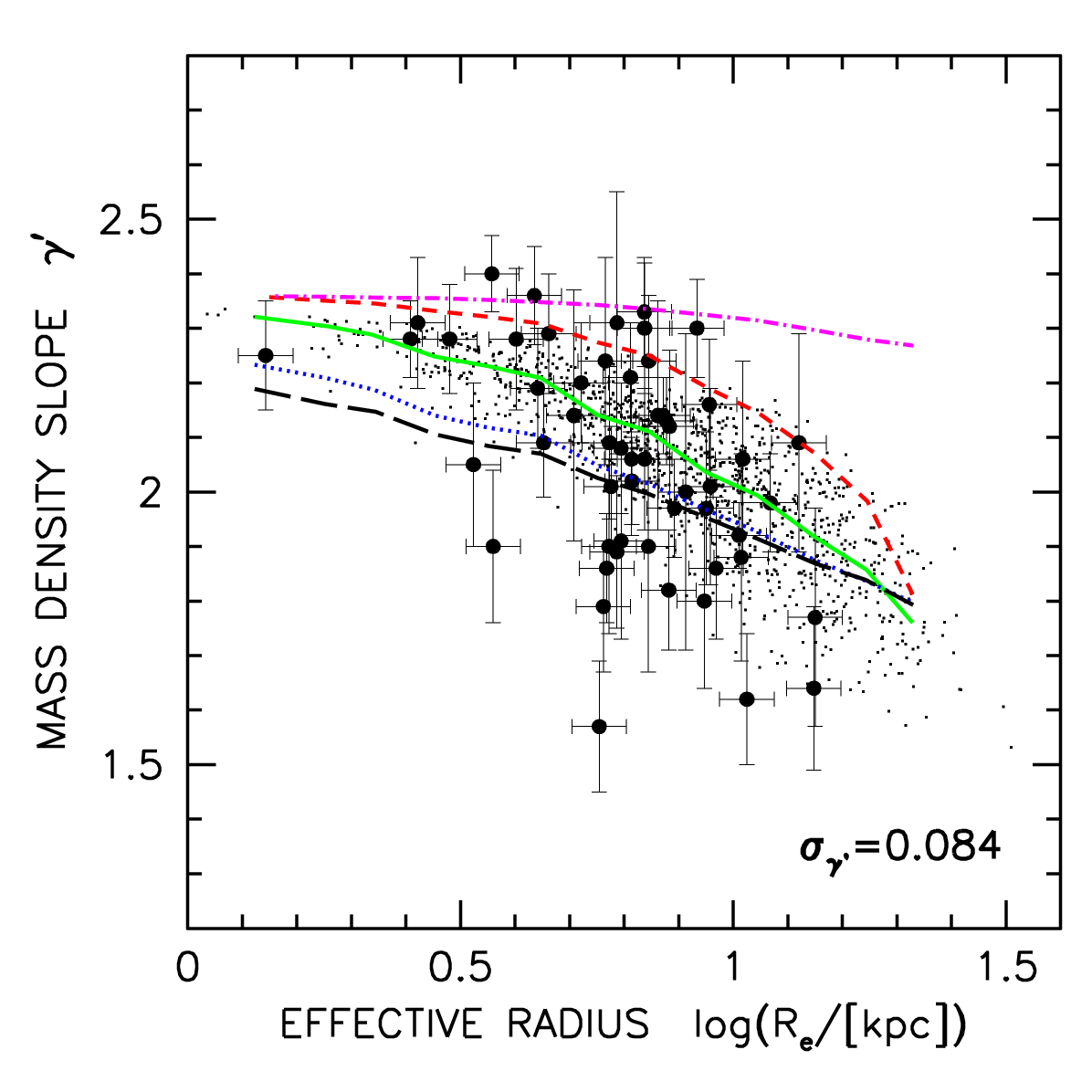,width=0.34\textwidth}
}
\centerline{
\psfig{figure=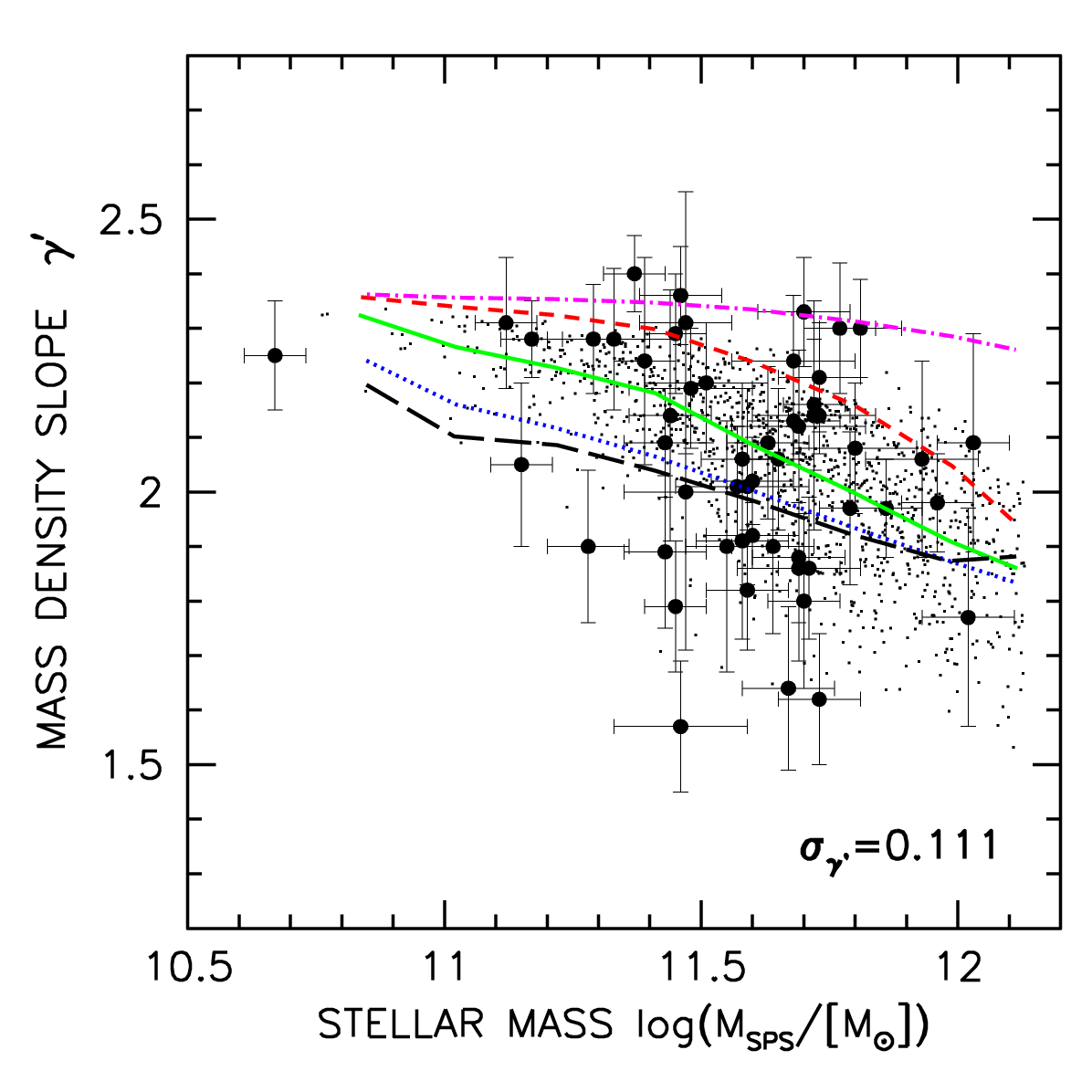,width=0.34\textwidth}
\psfig{figure=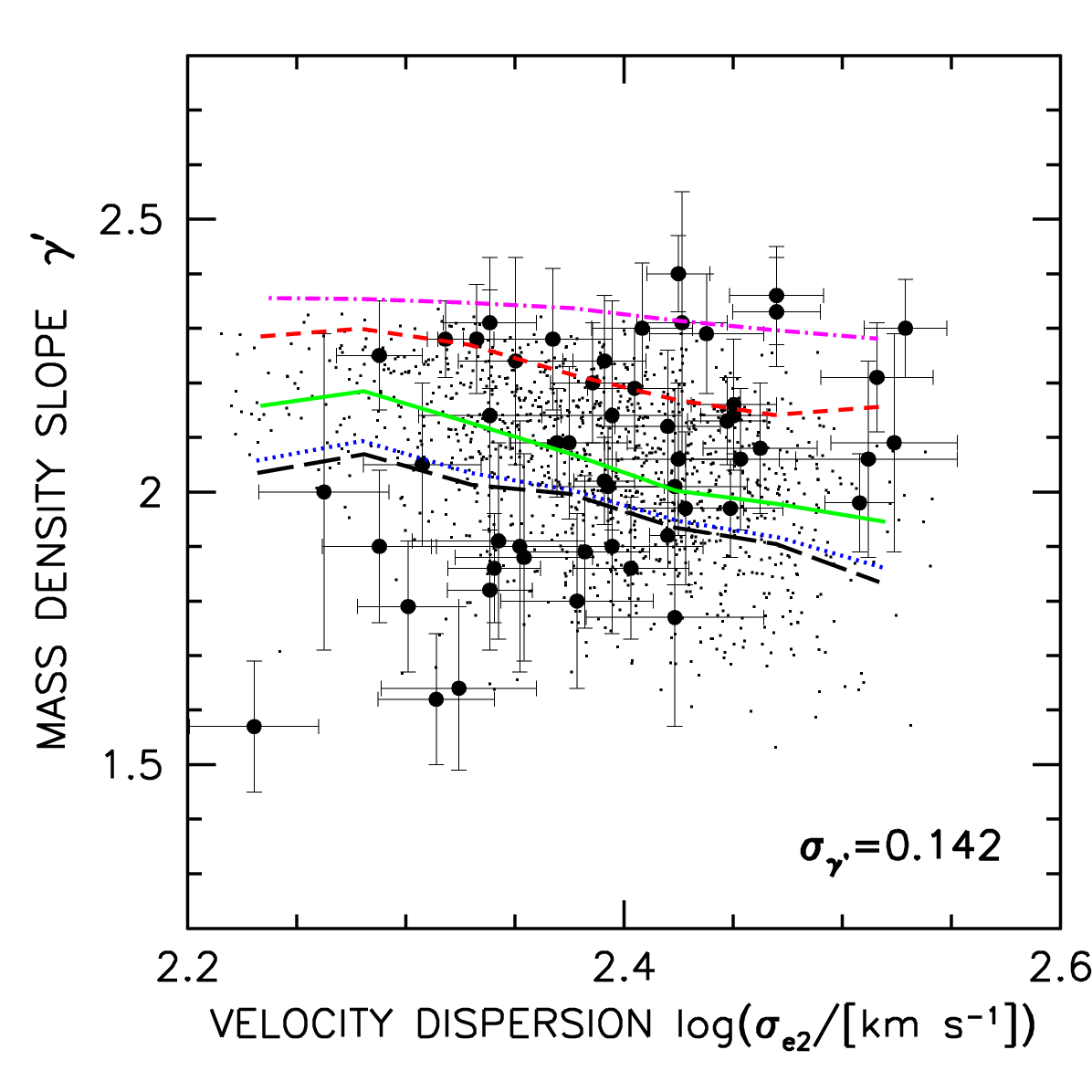,width=0.34\textwidth}
}
\caption{Correlations between mass density slope within the effective
  radius, $\gammap$, and galaxy observables: stellar surface density
  ($\Sigma_{\rm e}$, upper left), effective radius ($R_{\rm e}$, upper
  right), stellar mass ($M_{\rm SPS}$, lower left) and stellar
  velocity dispersion ($\sigma_{\rm e2}$, lower right). The models are
  given by coloured lines: adiabatic contraction ($\nu=1.0$, black
  long dashed); mild contraction ($\nu=0.5$, blue dotted); no
  contraction ($\nu=0.0$, green solid); halo expansion ($\nu=-0.5$,
  red short dashed). In addition, the points show individual galaxies
  for the no contraction model (which provides the best match to the
  observed mean and scatter of $\gammap$ -- see
  Fig.~\ref{fig:gamma-mean-sigma}). The scatter about the median
  relation for this model, $\sigma_{\gamma'}$, is given in the lower
  right corner of each panel. The tightest correlation in the models
  is between $\gammap$ and stellar density, while the weakest is with
  velocity dispersion.  The data from SLACS (Auger \etal 2010b) are
  given by filled symbols with error bars. For both models and data
  the stellar masses and surface densities have been scaled to a
  Salpeter IMF. The trends seen in the no contraction model are also
  seen in the observations, providing further support for its
  validity.}
\label{fig:gammae-Sigma}
\end{figure*}

Fig.~\ref{fig:gammae-Sigma} shows correlations between $\gammap$ and a
number of physical galaxy parameters: stellar surface mass density;
effective radius; stellar mass; and stellar velocity dispersion.
Lines show the median relations for the models, the points show the
model galaxies with uncontracted NFW haloes, and the filled points
with error bars show the observations from SLACS (Auger \etal
2010b). To ease comparisons between the different models and the data
the stellar masses and stellar densities have been re-scaled to a
Salpeter IMF. The most significant correlation in the data is
with stellar density, followed by effective radius and stellar
mass. These trends are qualitatively reproduced by the models.  To the
eye, the data appear to suggest a weak positive correlation with
velocity dispersion, which is of opposite sign to the models. However,
the measurement errors on $\gammap$ are positively correlated with the
errors on $\sigma_{\rm e2}$ (Auger \etal 2010b), and therefore the
trend is statistically insignificant (slope $0.07\pm0.08$) and
consistent with the weak opposite trend seen in our models.
 
Thus, not only does our model with uncontracted NFW haloes reproduce
the mean and scatter of the observed $\gammap$, it also reproduces the
correlations between $\gammap$ and other galaxy observables. We note
this achieved without any arbitrary fine-tuning of the distribution of
baryons and dark matter in galaxies.

\begin{figure*}
\centerline{
\psfig{figure=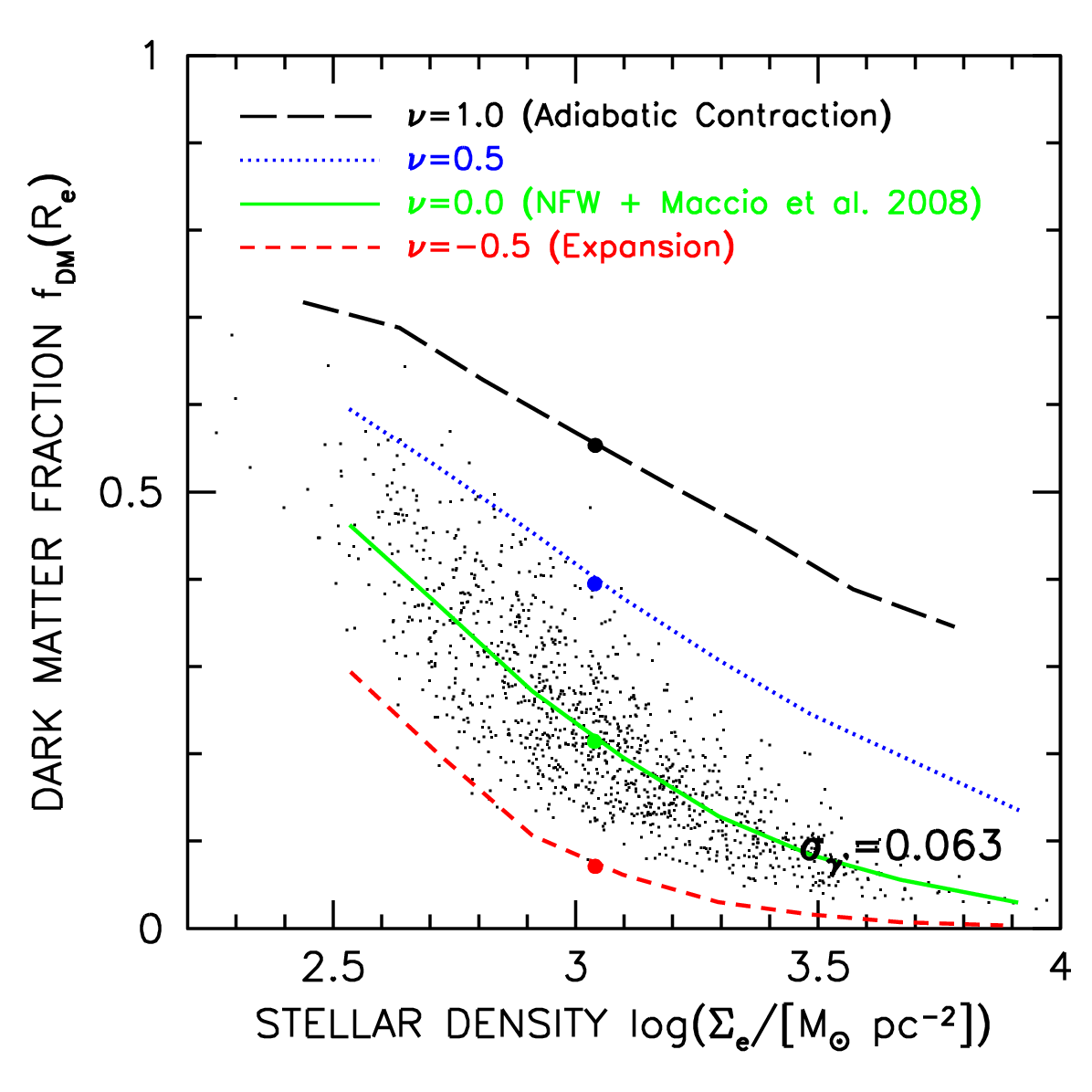,width=0.34\textwidth}
\psfig{figure=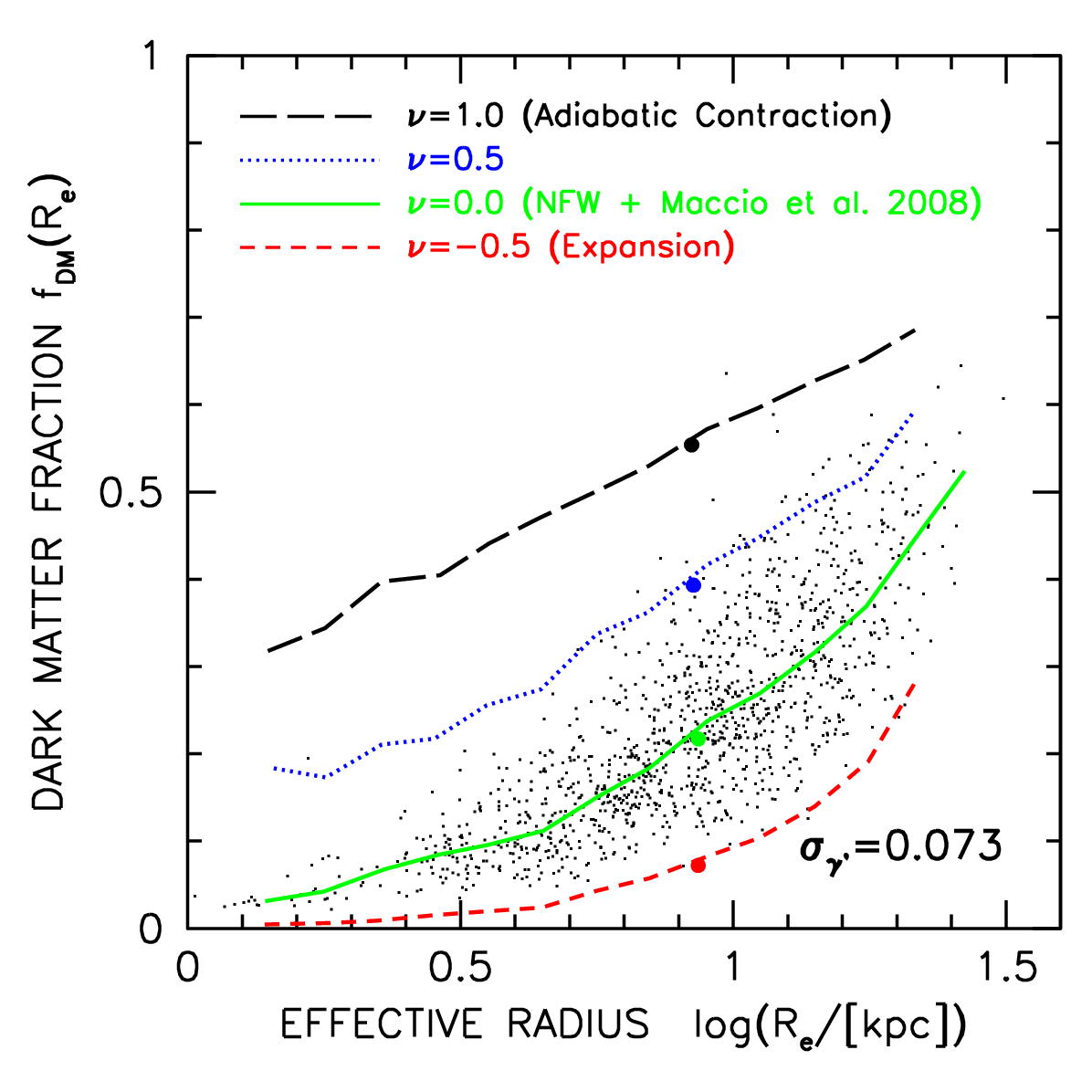,width=0.34\textwidth}
}
\centerline{
\psfig{figure=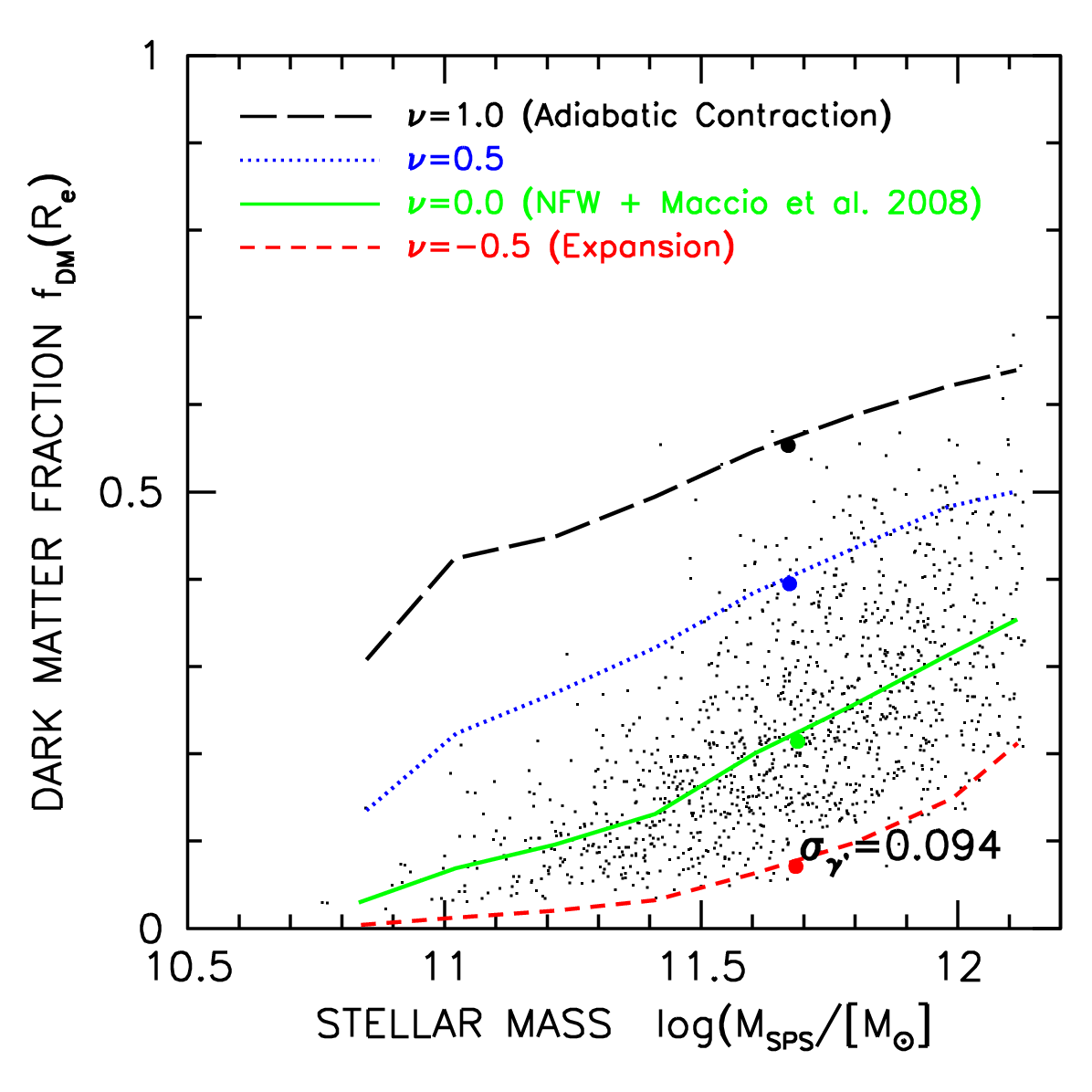,width=0.34\textwidth}
\psfig{figure=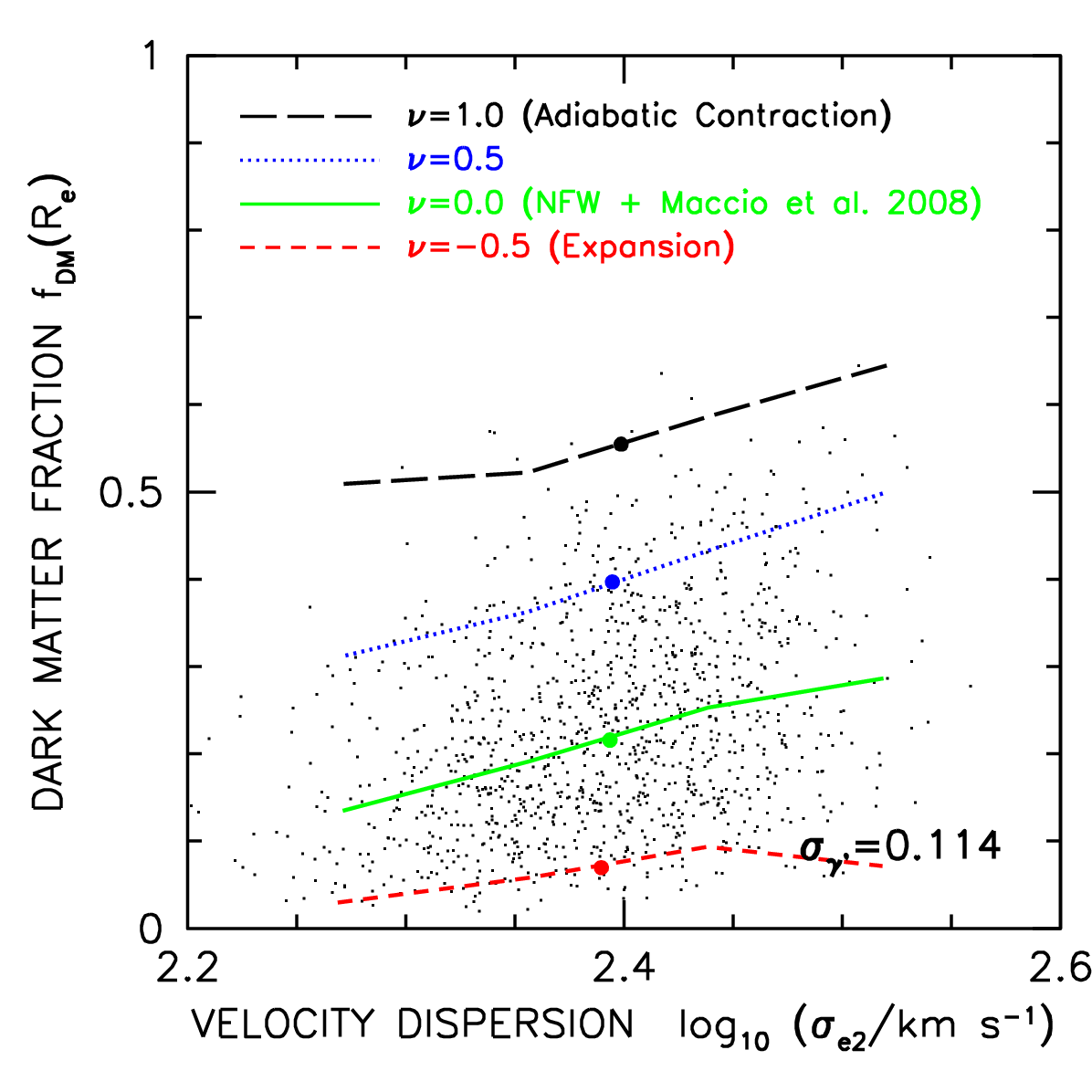,width=0.34\textwidth}
}
\caption{Correlations between dark matter fraction within the
  effective radius, $f_{\rm DM}(R_{\rm e})$, and galaxy observables
  (as in Fig.~\ref{fig:gammae-Sigma}) for our models. The scatter
  about the median relation for the no contraction model,
  $\sigma_{\gamma'}$, is given in the lower-right corner of each
  panel. The size of the scatter in these relations is directly
  correlated with the size of the scatter in the equivalent relations
  in Fig.~\ref{fig:gammae-Sigma}. }
\label{fig:fDM}
\end{figure*}

In our models, for a given halo response, there is almost no scatter
between $\gammap$ and dark matter fraction. Thus the strength of the
correlation between $\gammap$ and various galaxy parameters are
largely determined by the strength of the correlations with dark
matter fraction (see Fig.~\ref{fig:fDM}).

\section{Ensemble Mass density slope}
In this section, we comment on a method to infer the average mass
density slope of the dark matter within the effective radii of massive
early-type galaxies. This method was recently used by Grillo (2012) to
argue in favor of very steep dark matter density profiles, which is in
apparent contradiction to our results (as well as those by Auger \etal
2010a).

The main observables are the total projected mass, $M_{\rm tot}(R_{\rm
  Ein})$, within the Einstein radius, $R_{\rm Ein}$, obtained from
strong gravitational lensing.  For a given IMF one can calculate the
total mass in stars, $M_{\rm SPS}$, and the mass in stars within the
Einstein radius, $M_{\rm SPS}(<R_{\rm Ein})$.  Thus for each lens,
assuming an IMF, one can calculate the dark matter surface density
within the Einstein radius:
\begin{equation}
  \Sigma_{\rm DM}(R_{\rm Ein}) = \left[ M_{\rm tot}(<R_{\rm
    Ein}) - M_{\rm SPS}(<R_{\rm Ein})\right] /\pi R^2_{\rm Ein}.
\end{equation}
Since the Einstein radii are in general different for each lens
system, one can determine the average dark matter density profile for
an ensemble of gravitational lenses (see, e.g., Rusin \& Kochanek 2005
for a similar strategy).  However, the lenses have different masses
and sizes, and therefore the ensemble average cannot be conducted in
physical scales. A useful approach is to scale the size and density
measurements with the intrinsic scales of the galaxies: the effective
radii $\Re$, and effective surface densities $\Sigma_{\rm e}\equiv
{M_{\rm SPS}/R_{\rm e}^2}$. In this way, one can then define
dimensionless sizes and dark matter densities via
\begin{equation}
\Lambda = R_{\rm Ein}/R_{\rm e},
\end{equation}
and 
\begin{equation}
\Psi = \Sigma_{\rm DM}(R_{\rm Ein})/\Sigma_{\rm e}(R_{\rm e}).
\end{equation}
Plotting $\Psi$ versus $\Lambda$ thus yields an ensemble density
profile of the dark matter. The correlation can be described with a
power law:
\begin{equation}
  \Psi = \alpha \Lambda^{\beta}
\end{equation}
to yield an effective de-projected dark matter density slope
$\gammap_{\rm DM,g}=-\beta +1$.

Applying this method to 39 massive early-type lenses from the SLACS
survey, Grillo (2012) found $\beta=-1.04_{-0.22}^{+0.26}$
(corresponding to $\gammap_{\rm DM,g}=2.04^{+0.22}_{-0.26}$) for a
Chabrier IMF, and $\beta=-0.77_{-0.37}^{+0.62}$ (corresponding to
$\gammap_{\rm DM,g}=1.77^{+0.37}_{-0.62}$) for a Salpeter IMF. Since
an uncontracted NFW halo will have $\beta\simeq -0.1$ ($\gammap_{\rm
  DM}\simeq 1.1$), this provides marginal evidence for halo
contraction in response to galaxy formation (assuming the IMF is
lighter than Salpeter), in contradiction to our results. However, as
we discuss below, there are two issues that explain this apparent
discrepancy.

The first issue is that the IMF is likely Salpeter or heavier, which
will obviously require shallower dark matter slopes. For example, Auger
\etal (2010a), Dutton \etal (2013b), and Conroy \& van Dokkum (2012)
all favour IMFs Salpeter or heavier in the most massive galaxies. We
note that for a Salpeter IMF or heavier our results are in agreement
with Grillo's.

The second issue is more subtle and has to do with an implicit
assumption of the method used by Grillo (2012). Grillo assumes that
the re-scaling of the dark matter densities and Einstein radii does
not introduce spurious correlations. Specifically, in order for
$\Psi(\Lambda)$ to be equivalent to $\Sigma_{\rm DM}(\Rein)$ requires
that $\Sigma_{\rm SPS}/\Re^{\beta}$ is independent of $\Rein$. Using
SLACS data from Auger \etal (2010b), we find that this is not the case.
Fitting a relation of the form $\Sigma_{\rm e}/\Re^{\beta} \propto
(\Rein)^{\delta}$ on a grid of $\beta$ over the interval $[-2.5,1]$,
and then fitting a linear relation between $\delta$ and $\beta$ we
find that $\delta = -0.078 -0.586 \beta$.  Thus, the true dark matter
density slope is given by
\begin{equation}
\label{eq:correction}
 \gamma'_{\rm DM}\equiv \gamma'_{\rm DM, g} -\delta = 0.664 +0.414\gamma'_{\rm DM,g}.
\end{equation}
So a measured isothermal mass slope (i.e., $\gamma'_{\rm DM,g}=2$)
implies a shallower true mass slope of $\gamma'_{\rm DM}=1.49$.

Correcting the results from Grillo (2012) using
Eq.~\ref{eq:correction} yields $\gamma'_{\rm
  DM.g}=1.51_{-0.11}^{+0.10}$ for a Chabrier IMF, and $\gamma'_{\rm
  DM,g}=1.40_{-0.26}^{+0.15}$ for a Salpeter IMF.  In the context of
our models, this would favour weak contraction $(\nu\sim 0.5)$, which
is similar to that found in cosmological simulations by Abadi \etal
(2010). Allowing for an IMF slightly heavier than Salpeter (as we
favour) would increase the inferred $\beta$ to be in even closer
agreement with our results. In summary, we find that there is no
conflict between the ensemble average dark matter density slope as
derived by Grillo (2012) and our results based on the fundamental
plane and total mass density slopes

\section{Discussion}
\label{sec:dis}

We have shown that the observed distribution of mass density slopes
can be reproduced, {\it precisely}, if galaxies with realistic sizes
and concentrations are embedded in cosmologically motivated dark
matter haloes.  We note that a key requirement for our models to
reproduce the observed $\gammap$ is that the concentration of the
stars needs to be high enough so that $\gammap_{\rm star}(\Re) > 2$
--- because \LCDM haloes on their own never have such steep density
profiles on such scales. For example, galaxies with exponential
surface density profiles (i.e., characteristic of galaxy discs) would
be unable to result in $\gammap(\Re) \sim 2$.

Our model assumes that the scatter in galaxy sizes at fixed stellar
mass is uncorrelated with the structure of the dark matter halo. This
assumption was made for simplicity (and remarkably it seems to work),
but it should be tested using galaxy formation models. For example,
for spiral galaxies we {\it do} expect correlations between the
scatter in galaxy sizes with scatter in halo concentrations and galaxy
formation efficiencies (see Fig.7 in Dutton \etal 2007). It remains to
be seen whether such correlations exist for spheroid dominated
galaxies.

Our results favour unmodified cosmologically motivated (Navarro \etal
1997; Macci\`o \etal 2008) dark matter haloes. This is somewhat
surprising, since there are many processes that should modify the dark
halo structure (e.g., gas accretion, gas outflows, minor and major
mergers). One possibility is that these various processes are occurring
stochastically, and cancel each other out on average.  Alternatively,
we might expect that gas accretion dominates at early times, followed
by gas outflows, then dry mergers. Under this scenario, we would
expect the halo to initially contract, and then to slowly undo this
contraction over time. Thus a measurement of the evolution of the halo
response would be able to distinguish between these formation
scenarios for massive elliptical galaxies. This should be possible
with the current generation of galaxy scale lenses extending to higher
redshifts than SLACS (Sonnenfeld et al. 2013).

Recently, Chae \etal (2014) have performed a complimentary analysis of
the mass density profiles of early-type galaxies in the SDSS. These
authors find an average mass density slope of $<\gamma_{\rm
  e}>=2.15\pm0.04$, where individual $\gamma_{\rm e}$ are obtained
from power-law fits to the total density profiles between 0.1 and 1.0
effective radii. These results appear steeper than our uncontracted
halo models, as well as the slopes inferred from strong lensing
studies at low redshift (e.g., Auger \etal 2010b). However, since Chae
\etal (2013) adopt a different definition for the mass density slope, a
detailed comparison between our respective results is
non-trivial. Ideally, comparisons with mass density slopes derived
from strong lensing studies would employ the same method on model
galaxies as is applied to the lenses.


\section{Conclusions}
\label{sec:sum}

In the past few years observations have shown that massive early-type
galaxies have close to isothermal density profiles within one
effective radius (Koopmans \etal 2009; Auger \etal 2010b). We use
$\LCDM$-based mass models constructed to reproduce a number of the
observed scaling relations of early-type galaxies (including the
Faber-Jackson relation) to address two questions: (1) For models with
the correct distribution of baryons, do the properties of \LCDM haloes
result in models that reproduce the observed properties of $\gammap$?
(2) Do the observed properties of $\gammap$ help to distinguish between
models with different stellar initial mass functions?  We summarize
our results as follows.


\begin{itemize}

\item The median values of the mass density slopes of all our models
  are roughly isothermal: $1.8 \lta \gammap \lta 2.4$. The observed
  mean value of $\langle\gammap\rangle=2.08\pm0.03$ is best reproduced
  by a model with uncontracted NFW haloes and stellar masses $\simeq
  0.27$ dex higher than that obtained assuming a Chabrier (2003) IMF
  (Fig.~\ref{fig:gamma-mean-sigma}).

\item The scatter in $\gammap$ in our models with dark matter haloes
  is small $0.10 \lta \sigma_{\gammap} \lta 0.17$. The observed
  (intrinsic) scatter of $\sigma_{\gammap}=0.16\pm0.02$ is best
  reproduced by our model with uncontracted dark matter haloes
  (Fig.~\ref{fig:gamma-mean-sigma}).

\item In observations the mass density slope correlates with
  stellar surface density and effective radius, with higher $\gammap$
  in higher density and smaller galaxies. As with other properties of
  $\gammap$ this correlation is best reproduced by our model with
  uncontracted NFW haloes (Fig.~\ref{fig:gammae-Sigma}).

\item In the models the tightness of the correlations involving
  $\gammap$ are largely determined by the tightness of the
  corresponding correlations involving (spherical) dark matter
  fraction within the effective radius (Fig.~\ref{fig:fDM}).

\end{itemize}

In conclusion, our study shows that the many of the observed
properties -- including the bulge-halo conspiracy and other classic
scaling relations like the Faber Jackson and size mass relation -- of
massive early-type galaxies can be {\it precisely} reproduced by \LCDM
models under two conditions: (i) the IMF is not universal; (ii)
mechanisms, such as feedback from active galactic nuclei or dynamical
friction, effectively counterbalance on average the contraction of the
halo expected as a result of baryonic cooling.  We emphasize that no
correlations between the scatter in galaxy and dark halo parameters
(or equivalently, no fine-tuning between the baryons and dark matter)
is needed in order for our models to reproduce the observed small
scatter in total mass density profiles. Although our models are
clearly an oversimplified description of reality, they demonstrate
that self-consistent \LCDM-inspired models can be found. This gives
hope that the recent improvements in numerical simulations including
baryonic physics could soon lead to realistic models for the formation
of massive type galaxies in quantitative agreement with the tight
constraints provided by observations. In turn, this might help explain
what is the relative contribution of the different processes that
contribute to produce the effective profiles that we observe.
Our work also provides further evidence against a universal IMF for
galaxies. In agreement with earlier studies (Auger \etal 2010a; van
Dokkum \& Conroy 2010; Spiniello \etal 2011; Sonnenfeld \etal 2012,
Conroy \& van Dokkum 2012; Smith \etal 2012; Spiniello \etal 2012;
Cappellari \etal 2013; Dutton \etal. 2013a,b), we find that a ``heavy''
Salpeter-like IMF is preferred for massive galaxies over the light
Chabrier-like IMFs usually preferred for Milky Way-type galaxies. The
origin of this non-universality is still hotly debated. However, the
fact that the only quantity that seems to correlate with the mass
structure of massive early-type galaxies is the central surface
density could be a clue that the density of stars (and therefore
possibly gas at the epoch of peak star formation) is an important
ingredient of shaping the final stellar initial mass function (Hopkins
2013).

\section*{Acknowledgements}
We thank our friends and colleagues of the SLACS and SWELLS
collaborations for numerous stimulating discussions about these topics
over the years. We also thank the participants of Lorentz Workshop
``Is the stellar IMF universal?'', in particular the organizers (Leon
Koopmans, Scott Trager, Romeel Dav\'e, Patrick Hennebelle, Chris
McKee, Michael Meyer, Stella Offner) for illuminating presentations
and conversations. We are also grateful to the staff at Lorentz Center
in particular Ms Gerda Filippo for her making the workshop a wonderful
experience.

AAD acknowledges financial support from the National Science
Foundation Science and Technology Center CfAO, managed by UC Santa
Cruz under cooperative agreement no. AST-9876783. 
TT acknowledges support from the NSF through CAREER award NSF-0642621,
and from the Packard Foundation through a Packard Research Fellowship.
This research was partially supported by NASA through Hubble Space
Telescope programs GO-10587, GO-10886, GO-10174, 10494, 10798, 11202.


\label{lastpage}
\end{document}